\documentclass[preprint]{elsart}
\usepackage{amsmath,amssymb,epsfig,color}
\def\be{\begin{eqnarray}}
\def\ee{\end{eqnarray}}
\newcommand{ \al}{$ \alpha$~}
\newcommand{\psid}{\psi^{\dagger}}
\newcommand{\vecx}{\mathbf x}

\newcommand{\vecnabla}{\mathbf \nabla}

\begin{document}
\begin{frontmatter}
\title{Alpha matter on a lattice}
\author{Armen Sedrakian$^1$, Herbert M\"uther$^1$ and Peter Schuck$^2$}
\address{$^1$Institut f\"ur Theoretische Physik,
 Universit\"at T\"ubingen,\\ 72076 T\"ubingen, Germany\\
$^2$Groupe de Physique Th\'eorique, Institut de Physique Nucl\'eaire,\\
            91406 Orsay, France
}
\date{\today}
\begin{abstract}
We obtain the equation of state of interacting alpha matter and the
critical temperature of Bose-Einstein condensation of alpha particles
within an effective scalar field theory. We start from a
non-relativistic model of uniform alpha matter interacting
with attractive two-body and repulsive three-body potentials and
reformulate this model as a $O(2)$ symmetric scalar $\phi^6$
field theory with negative quartic and positive sextic interactions.
Upon restricting the Matsubara sums, near the temperature
of Bose-Einstein condensation, to the zeroth order modes
we further obtain an effective classical
theory in three spatial dimensions. The phase diagram of the alpha
matter is obtained from simulations of this effective field theory
on a lattice using local Monte-Carlo algorithms.
\end{abstract}

{\tt \small
}                        %  thermodynamical, statistical and structural
                         %  properties.
\\
\end{frontmatter}
\newpage
%------------- Introduction/motivation ---------------------
\section{Introduction}
The  \al  particle model of nuclei has its origin in the Gamow's
theory of nuclear \al emission,
which assumes pre-formed  \al  particles inside a nucleus.
Light self-conjugate $4N$ ($N=1,2,\dots$) nuclei, called also   \al  nuclei,
have been associated with the
 \al  particle model, which assumes  that  \al  nuclei are
composed of structureless rigid  \al  particles as their
stable constituents;  \al  itself is an extremely stable nucleus, and is
emitted by naturally radioactive nuclei and the  \al  nuclei
which are more stable than their neighbors \cite{BLATT52}.
This model in its entirety or some of its basic
assumptions has been used  over years
in the  \al  cluster structure investigations \cite{ALPHA_STRUC}. The
 \al  particle models are more successful in explaining the excited
states of  \al  nuclei in terms of their rotational states than their
ground state properties;  \al  clustering appears as  one of the
many transient types of clusterings that are continuously
emerging and disappearing inside a nucleus \cite{NOTE1}.

The large $N$ limit of \al particle model describes
{\it \al matter} -  an extended uniform system of structureless,
indistinguishable  \al  particles in a uniform background
of neutralizing charge. Alphas maintain their integrity at densities where the
inter-alpha distance is much larger than the average size of an alpha
particle ($\le 2$ fm) and at temperatures not too large compared
to the binding energy per particle. Deconfined alpha matter occurs
in astrophysical settings. At low densities/temperatures,
characteristic for the nucleosynthesis site in supernova
debris, the neutron to proton ratio and the ultimate outcome of
the $r$-process nucleosynthesis depends on the
recombination kinetics of nucleons into \al particles \cite{MCLAUGLIN}. At
higher densities (but still below the nuclear saturation density)
a substantial fraction of hot and dense supernova matter
resides in \al particles~\cite{LAMB,LLPR,ROEPKE_EOS,LS}. The
equation of state of subnuclear matter~\cite{LLPR,LS}
is at the heart of the simulations of the supernova collapse
and explosions and has important implications for the formation
of the supernova neutrino signal.

The studies of \al matter isolated from the astrophysical
environment it is embedded in provide a first approximation
for accessing the role of correlations, phase transitions and
condensation phenomena within this component. There have 
been many detailed studies of the ground state of uniform \al
matter~\cite{TAMAGAKI,CLARK1,CLARK2,CLARK3,CLARK4,BRINK,TURNER,AGUILERA,THOSAKI},
including the  benchmark variational calculations of Clark and
co-workers \cite{CLARK1,CLARK2,CLARK3,CLARK4} to which we compare our results.
Alpha matter in its ground state may form a bcc lattice,
which takes an advantage of closest possible packing of particles without
destroying their identity \cite{BRINK,AGUILERA,THOSAKI}.
Recent work focused on the possibility of the Bose-Einstein condensation
of  \al  particles in  \al  nuclei \cite{SCHUCK1,SUZUKI,SCHUCK2} and the
competition between two- and four-body clustering in
infinite matter~\cite{ROEPKE}.

The aim of this work is the study of uniform \al matter from the
lattice field theory point of view. First, we show  that the
symmetric (non-condensate) phase of \al matter interacting
with attractive two-body and repulsive three-body forces,
belongs to the universality class of $O(2)$ symmetric  scalar
$\phi^6$ field theories. Second, we simulate the scalar
$\phi^6$ field theory on a lattice using Monte-Carlo
algorithms. The parameters of the continuum action are deduced
from the non-relativistic interactions derived from the fits
to the  \al-\al  scattering. For several parameter sets
we then obtain the equation of state of \al matter.
Our approach is complementary to the other methods
to study the \al matter; an advantage of the present approach is the
close connection to the field of critical phenomena, where
many universal properties  which are independent of
details of interactions in the system (e.g. critical exponents)
are known.

Our work is in part motivated by the studies of the scalar $\phi^4$
field theory which describe the shift in the critical
temperature of the Bose-Einstein
condensation for repulsive atomic systems
\cite{BAYM,ATOMBEC1,ATOMBEC2,ATOMBEC3,ATOMBEC4,ATOMBEC5}.
These studies are suited for treating repulsive, dilute Bose gases
in the vicinity of the critical temperature of
Bose Einstein condensation. Alpha matter is attractive at large
and repulsive at small distances. To describe these features,
we adopt the scalar $\phi^6$ theory with negative quartic and
positive sextic couplings. While we concentrate here on alpha
matter we anticipate that our model could be applied to dilute
Bose gases interacting with  long-range attractive
and short range repulsive interactions.
Section~II derives the effective action of the classical
three-dimensional theory starting from the Hamiltonian of \al-matter
interacting with two- and three-body forces. In Section~III we
describe the Monte-Carlo simulations of this theory on a lattice.
The results for the equation of state of \al-matter and the critical
temperature of Bose-Einstein condensation are presented in Section~IV.
Section~V is a short summary of this work.

%----------------------- Basic set-up ------------------------
\section{Effective action}
A uniform, non-relativistic \al matter is described by the Hamiltonian
\be\label{eq:1}
H &=& \int\!d^3x \Biggl[ \frac{\hbar^2}{2m}
     \mathbf\nabla\psid(\vecx)\mathbf\nabla\psi(\vecx)
-    \mu\psid(\vecx)\psi(\vecx)
%\nonumber\\&&
+\psid(\vecx)\psi(\vecx)U(\vecx )\Biggr],\nonumber\\
\ee
where $m$ is the \al particle mass, $\mu$ is the chemical potential,
$\psi(\vecx)$ and $\psid(\vecx)$ are the creation and
annihilation operators, and
\be\label{eq:2}
U(\vecx) &=& \int\!d\vecx' V_2(\vecx',\vecx)
     \psid(\vecx')\psi(\vecx')\nonumber\\
&+& \int\!d\vecx'\int\!d\vecx'' V_3( \vecx,\vecx',\vecx'')
     \psid(\vecx'')\psi(\vecx')\psid(\vecx'')\psi(\vecx'), %\nonumber\\
\ee
where $V_2$ and $V_3$ are the two- and three-body interaction
potentials (which are not identical to the free-space 
interactions in general, see below).
To reformulate the theory above as an effective field
theory, we replace the finite range potentials by zero range
constant couplings (which are constrained to reproduce
the low-energy scattering data) and adopt the lattice regularization.
For contact form of interaction Eq. (\ref{eq:2}) reduces to
\be\label{eq:3}
U(\vecx) &=&  g_2\psid(\vecx)\psi(\vecx)
  +  g_3\psid(\vecx)\psi(\vecx)\psid(\vecx)\psi(\vecx),
\ee
where $g_2 = 4\pi\hbar^2a_{\rm sc}^{(2)}/m$ and the two-body
scattering length is defined as
\be \label{ASCAT}
 a_{\rm sc}^{(2)} = \frac{m}{4\pi\hbar^2}\int d\vecx
~V_2(\vecx).
\ee
For systems featuring short-range repulsive interactions (hard core)
the two-body potential in Eq. (\ref{ASCAT}) should be replaced by
the scattering $T$-matrix which sums up the particle-particle ladder
series. We follow the convention  that assigns postive sign to the
scattering length for attractive interactions.  In the case, where
the three-body interactions are represented by a sum of pairwise (repulsive)
two-body interactions:
$V_3( \vecx,\vecx',\vecx'')
=\tilde V_2^{(1)}( \vecx-\vecx')
+\tilde V_2^{(2)}(\vecx-\vecx'')
+\tilde V_2^{(3)}( \vecx'-\vecx'')
$ \cite{KATO}, the three-body coupling constant is
$g_3 = 4\pi\hbar^2\tilde a_{\rm sc}^{(3)}/3m$, where
$a_{\rm sc}^{(3)}$ is defined via a spatial integral (\ref{ASCAT})
over the two-body potential $\tilde V_2^{(i)}(\vecx-\vecx')$.

To map the theory defined by Eq.~(\ref{eq:1}) on an effective 
scalar field theory,  we adopt the finite temperature Matsubara 
formalism, described in refs. \cite{BAYM,ZINN}.
Consider the fields $\psi$ and $\psi^{\dagger}$ as periodic 
functions of the imaginary time $\tau\in [-\beta, \beta]$, where
$\beta = 1/T$ is the inverse temperature. 
These fields can be decomposed into discrete Fourier series
\be\label{FOURIER}
\psi(\vecx,\omega_{\nu}) &=& \sum_{\nu=-\infty}^{\infty}e^{i\omega_{\nu}\tau}\psi(\vecx,\tau)
\nonumber\\
 &=& \psi_0(\vecx)
 +\sum_{\nu=-\infty,~\nu\neq 0}^{\infty}e^{i\omega_{\nu}\tau}\psi(\vecx,\tau),
\ee
where the Fourier frequencies $\omega_{\nu}$ are the bosonic Matsubara
modes $\omega_{\nu} = 2\pi i \nu T$ ($\nu$ assumes integer values).
The Matsubara Green's function for our system can be written as 
\be 
{G}^M(\omega_{\nu},\vecx) = [i\omega_{\nu} -(2m)^{-1}\vecnabla^2
+\mu]^{-1},
\ee
where the chemical potential includes contribution from the self-energy. 
To clarify this point, consider the spectrum 
of excitations of the system which is determined above the 
critical temperature $T_c$ by the poles
of the retarded propagator. This function is obtained from the Matsubara 
Greens function via the replacement $i\omega_{\nu}\to \omega +i\eta$
(where $\omega$ is real)
\be 
{G}^R(\omega,\vecx) = [\omega -(2m)^{-1}\vecnabla^2  - {\rm Re}
\Sigma^R(\vecx, \omega) +\mu_f + i\eta]^{-1},
\ee
where $\mu_f$ is the chemical potential of non-interacting gas and 
$\Sigma^R(\vecx, \omega)$ is the retarded self-energy. The 
contribution of the two-body interaction to the retarded
self-energy can be written as
\be 
\Sigma_{(2)}^R(\vecx, \omega)  = \int d\vecx'd\omega'\Gamma_{(4)}(\vecx,\vecx',\omega+\omega') 
{\rm Im}G^R(\vecx',\omega'), 
\ee
where $\Gamma_{(4)}$ is the four-point vertex, which to the lowest order 
reduces to the potential $V_2$, but in general may
include two particle resummation series. The three-body contribution to the self-energy, which results from the three-body repulsive interaction, 
can be written in terms of six-point vertex 
$\Gamma_{(6)}$ in similar fashion. For contact interactions 
$\Gamma_{(4)} = g_2$ and $\Gamma_{(6)} = g_3$ and the net 
self-energy is a constant $\Sigma_0$. Thus the retarded propagator can be written 
as 
\be \label{GR}
{G}^R(\omega,\vecx) = [\omega -(2m)^{-1}\vecnabla^2 +\mu + i\eta]^{-1},\quad \mu = \mu_f - \Sigma_0.
\ee
For temperatures $ T\geq T_c$  a macroscopic number of particles 
starts  to occupy the zero-momentum state. Since the chemical 
potential must stay below the lowest possible energy 
state to keep the occupation probabilities positive, it 
follows from Eq. (\ref{GR}) that $\mu <0$; because a 
macroscopic  number of particles occupies the 
zero-momentum ground state  as $T\to T_c$ we need $\mu \to 0$. 
Since the above argument is equivalent to the statement that the fugacity 
\be 
z = e^{\beta\mu} \to 1,
\ee
the characteristic scales of spatial variations of the Green's function 
with non-zero Matsubara frequencies are $l = (2m\omega_{\nu})^{-1/2}$, 
which are of the order of the thermal wave-length $\lambda = 
(2\pi/mT)^{1/2}$. The contribution of the non-zero Matsubara modes to 
the sum in Eq. (\ref{FOURIER}) can be neglected if one is interested 
in the scales $L\gg l$, which are characterized by the zero-frequency 
modes.
     
Upon dropping the second term in Eq. (\ref{FOURIER}) and
introducing two new real scalar fields $\phi_1$ and $\phi_2$
\be \label{eq:4}
\psi_0 = \eta(\phi_1+i\phi_2),
\quad
\psid_0 = \eta(\phi_1-i\phi_2),
\ee
where $\eta = \sqrt{m/\hbar^2\beta}$, the continuum action
takes the form
\be \label{eq:6}
&&{S}\left(\phi\right)
 = \int d^3 x   \Biggl\{
\frac{1}{2}\sum_{\nu}\left[\partial_{\nu}\phi(\vecx)\right]^2
\nonumber\\
&&\hspace{4cm}+\frac{r}{2}\phi(\vecx)^2
-\frac{u}{4!}\left[\phi(\vecx)^2\right]^2
+\frac{w}{6!}\left[\phi(\vecx)^2\right]^3
\Biggr\},
\ee
where
$\phi^2 = \phi_1^2+\phi_2^2$, $r = -2 \beta\mu\eta^2$,
$u = 4! \beta g_2 \eta^4$, $w = 6!\beta g_3 \eta^6$.
The action (\ref{eq:6}) describes a classical
$O(2)$ symmetric scalar $\phi^6$ field theory in three spatial dimensions.
Note that the positive sextic interaction guarantees that the
energy is bound from below, which otherwise would be unbound
because of the negative sign of the quartic term that
describes the attractive two-body interactions.
The characteristic length scale of the theory is set
by the parameter $u$, which has dimension of inverse length;
the dimensionless parameter of the lattice theory
is $ua_L$, where $a_L$ is the lattice spacing.
The theory described by the action (\ref{eq:6}) differs from  the ordinary
$\phi^4$ theory by the order of the phase transition it predicts.
Indeed, the $\phi^6$ theory admits a first order phase transition  since
at low temperatures the free energy develops a second minimum
with the field $\phi$ assuming two equilibrium values, one at the origin
$\langle \phi\rangle = 0$  and a second along the circle of constant
radius$\sqrt{ \langle\phi_1^2\rangle+\langle\phi_2^2\rangle}$.
While the first order (liquid-gas) phase transition in
\al matter is of interest in its own right, below we concentrate on a
the possibility of the Bose-Einstein condensation of \al particles.
This phenomenon
is characterized by a pole in the distribution of Bose particles,
which then occupy the zero momentum ground state of the system - a state
that is characterized by the condition $\mu \to 0$
at fixed temperature and density.

The thermodynamic functions of the model are obtained from the
partition function
\be \label{eq:5}
{Z} = \int [d\phi(\vecx)] {\rm exp}
\left[-{S}\left(\phi\right)\right].
\ee
For example, the expectation value of the particle number density
$n_{\alpha} = \langle \psi^*\psi\rangle = (\beta V)^{-1}
\partial {\rm ln}{Z} /\partial \mu$, where $V$ is the volume.

The coupling constants of the continuum theory
can be related to the  values of the two-body scattering length
$a_{\rm sc}$ (derived from  \al-\al  potentials fitted
to the scattering data) only in the dilute limit.
The details of short range physics become unimportant
when the condition $\lambda \gg a_{\rm sc}$ is fullfiled, where
$\lambda^2 =2\pi/\eta^2$ is the thermal wavelength
of \al particles.
However, the concept of scattering length can be used for  dense
systems if it is defined in terms of the full scattering
$T$-matrix, instead of the Born amplitude (see e.g. \cite{LP}).
Since the two-body  \al-\al  potentials are characterized by hard cores
(strong, short-distance repulsion) we shall treat the scattering length as a
parameter with the understanding that it can be derived from an
appropriate $T$-matrix (or phase-shifts), where the effects of the
hard-core are removed by a full resummation of the ladder series.

\section{Numerical simulations}

For the purpose of numerical simulations we discretize the continuum
theory on a lattice by replacing the integrations over a summation
over the lattice sites. The discretized version of
the continuum action (\ref{eq:6}) is
\be \label{eq:7}
{S}_L\left(\phi\right) &=& \sum_i \Biggl\{
-2\kappa \sum_{\nu}\phi_L(\vecx)\phi_L(\vecx+a\hat\nu)
\phi_L(\vecx)^2 + \lambda
\left[1+\phi_L(\vecx)^2\right]^2\nonumber\\&-&
 \lambda+ \zeta\left[\phi_L(\vecx)^2\right]^3
\Biggr\},%\nonumber\\
\ee
where the $i$-summation is over the sites of a 3d cubic lattice,
the $\nu$ summation is over unit vectors in three spatial direction
(nearest neighbor summation).
The hopping parameter $\kappa $, the two- and three-body
coupling constants $\lambda$ and $\zeta$ are related to the
parameters of the continuum action
through the relations $a_L^2r = (1-2\lambda)/\kappa-6$,
$\lambda = a_L\kappa^2 u/6$ and $\zeta = w\kappa^3/90$.
The lattice and continuum fields are related by $\phi_L(\vecx)
= (2\kappa/a_L)^{1/2}\phi(\vecx)$.
The components of the spatial vector $x_{\nu}$ assume integer numbers of
the lattice spacing $a_L$: $x_{\mu} = 0, a_L, \dots (L_{\nu}-1)a_L$.
In our numerical simulations we adopt a simple cubic lattice in
three spatial dimensions with periodic boundary
conditions imposed on the field variable: $\phi_L(x+a_LL)
= \phi_L(x)$  [for a box of length $L$
there are $L^3$ (real) variables within the volume $(La_L)^3$].

The lattice regularization of a field theory provides an effective
cut-off $\Lambda\sim a_L^{-1}$ and breaks its
translational and rotational invariance. To restore the
invariance and recover the continuum  theory the limit
$ua_L\to 0$ should be taken in the numerical data. In the context
of $\phi^4$ theory in four (and higher) spacetime dimensions it has
been proven that the continuum limit of the theory is trivial, i.e. the
renormalized couplings vanish in the continuum limit (this would not be
inconsistent for an asymptotically free theory). In three
dimensions the $\phi^4$ and $\phi^6$ theories are
superrenormalizable. We leave the
study of the renormalization group equations governing the present 
$\phi^6$ theory for the future. 
Here we take the point of view of the effective
field theory, which exploits the fact that the cut-off can be made so
large (spacing so small) that the cut-off effect in the
Green's functions of the theory are completely negligible
\cite{PARISI}. In other words,
we require the correlation length $\xi$ of the theory to satisfy
the condition $\xi \gg a_L$. The correlation length is deduced by
computing the correlation functions on the lattice
$\langle \phi(x)\phi(y)\rangle$ for a fixed lattice spacing $a_L$. In the
simulations described below it was found that fixing the
physical lattice spacing $a_L = 1$ fm guarantees that the condition
$\xi \gg a_L$ is fulfilled; this would correspond to introducing a hard
core of a radius $\sim 1$ fm in the interactions, which is consistent
with the size of the cores of  \al-\al  potentials.

Another limit that must be taken in the lattice Monte-Carlo simulations
is the limit $uL\to \infty$ to minimize the finite lattice size effects;
in other words,
the thermodynamic parameters of the system in a finite box must
have a limit when the size of the box goes to infinity, and this
limit should not depend on the periodic boundary conditions imposed.
In practice we successively increase the size of the
lattice and compare the observables (e.g. the expectation values
of the fields and the correlations functions) until a convergence
is reached. The finite size effects become insignificant when
the condition $L\gg\xi $ is fulfilled.

Within the effective field theory approach sketched above the
``bare'' parameters $r$, $u$, and $w$ can be used to derive the
parameters of the lattice action for a specific choice of lattice
spacing $a_L$. We note here that the parameters of the continuum
action themselves are renormalized due to the higher order Matsubara
modes. These modifications, which are expected to be small in the
vicinity of the critical temperature of phase transition, are
neglected.

A universal feature of the potential models of the nuclear
\al-\al  interaction is a repulsion for small separations, which is
weakening as the relative orbital angular momentum $l$ increases from
0 to 4, followed by an attractive tail that is largely independent of
$l$. The range of the inner repulsion is approximately 2 fm, the range of
the outer attraction is about  5 fm.
For the hard-core plus a square well potential of Van der Spuy and
Pienaar (VP) \cite{VP}, with the core radius
$a = 1.7$~fm, the well depth $V_0=7.2$~MeV
and well width 4~fm, one finds the $S$ wave scattering length
$a_{\rm sc} = -7.33$~fm which translates into a coupling constant
$g_2 = -954.3$ MeV fm$^3$. Below we shall vary the magnitude of
$a_{\rm sc}$ to account for the renormalization by the
short range effects and differences among the available
\al-\al  interactions. For the three
body coupling constant we employ the potential of
Ref. \cite{KATO} to obtain $g_3 = 5944.8$  MeV fm$^3$. Table 1 lists
the parameters of the continuum and the lattice theories which
were used in the numerical simulations
for $a_L = 1$~fm.
\begin{figure}[t] % fig 1
\psfig{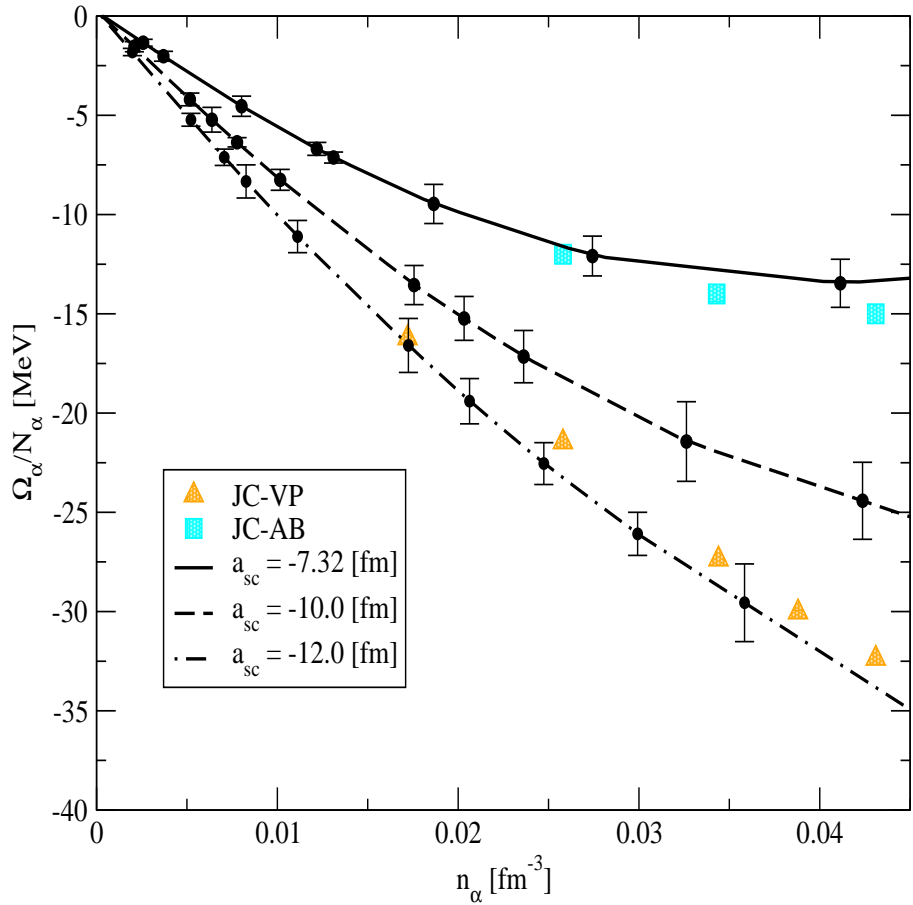}
\caption{%(Color online)
 Dependence of the energy (per particle) of the  \al  matter
 on the density  for several values of the scattering length at
 constant fugacity ${\rm log} ~z = -0.1$. The squares
 and the triangles show  the variational results of
 Johnson and Clark \cite{CLARK4} for the potentials of ref. \cite{VP}
 and \cite{AB}, respectively.
}
\label{MSfig:fig1}
\end{figure}
\begin{figure}[htb] % fig 2
\psfig{figure=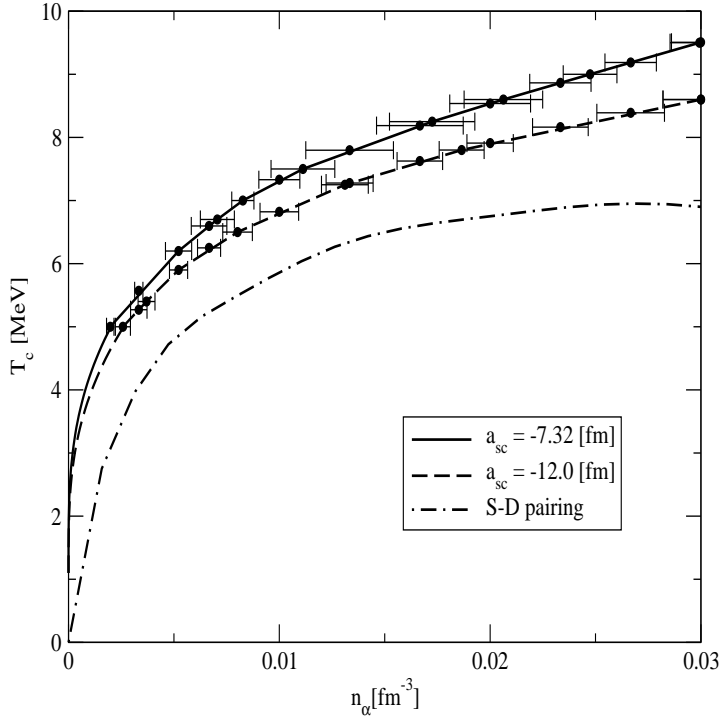,height=14cm,width=12.cm,angle=-90}
\caption{%(Color online)
The density dependence of the critical temperature
of $ \alpha$ condensation for scattering lengths
$a_{\rm sc} = 12$ fm (triangles) and  $a_{\rm sc} = 7$ fm (squares).
The critical temperature for isoscalar quasideuteron pairing
\cite{LOMBARDO} is shown by dashed dotted line.
}
\label{MSfig:fig2}
\end{figure}

\begin{table}
\caption{The  parameters of the continuum and lattice theories
for several values of the temperature  and
two-body scattering lengths at constant  ${\rm log}~ z= -0.1$.
The parameters of the lattice theory are listed for $a_L = 1$ fm.
}
\begin{tabular}{ccccccc}
\hline
$T$ &  $r$  [fm$^{-2}$]  & $u/10^3$  [fm$^{-1}$] & $w/10^5$
& $\kappa$ & $\lambda$ & $\zeta$ \\
\hline
&&$a_{\rm sc}^{(2)}$ = -7.33 fm &\\
\hline
9.0 & 0.174 & 0.959 & 3.113 & 0.047 & -0.355 & 0.361\\
8.0 & 0.154 & 0.853 & 2.460 & 0.049 & -0.348 & 0.331\\
7.0 & 0.135 & 0.746 & 1.883 & 0.052 & -0.340 & 0.299\\
6.0 & 0.116 & 0.640 & 1.384 & 0.056 & -0.330 & 0.265\\
5.0 & 0.096 & 0.533 & 0.961 & 0.060 & -0.318 & 0.228\\
\hline
&&$a_{\rm sc}^{(2)}$ = -10.0 fm &\\
\hline
9.0 & 0.174 & 1.309 & 3.113 & 0.041 & -0.372 & 0.244\\
8.0 & 0.154 & 1.164 & 2.460 & 0.043 & -0.366 & 0.224\\
7.0 & 0.135 & 1.018 & 1.883 & 0.046 & -0.359 & 0.204\\
6.0 & 0.116 & 0.873 & 1.384 & 0.049 & -0.350 & 0.181\\
5.0 & 0.096 & 0.727 & 0.961 & 0.053 & -0.339 & 0.158\\
5.00 & 0.10 & 0.73 & 0.96 & 0.053 & -0.339 & 0.158\\
\hline
&&$a_{\rm sc}^{(2)}$ = -12.0 fm &\\
\hline
9.0 & 0.174 & 1.571 & 3.113 & 0.038 & -0.382 & 0.193\\
8.0 & 0.154 & 1.397 & 2.460 & 0.040 & -0.376 & 0.178\\
7.0 & 0.135 & 1.222 & 1.883 & 0.043 & -0.369 & 0.162\\
6.0 & 0.116 & 1.048 & 1.384 & 0.045 & -0.361 & 0.145\\
5.0 & 0.096 & 0.873 & 0.961 & 0.049 & -0.350 & 0.126\\
\hline
\end{tabular}
\end{table}

\section{Results}

The field configurations on the lattice were evolved using a
combination of the heatbath and local Metropolis algorithms
with the number of equilibration sweeps $10^5-10^6$.
Lattices with sizes from $8^3$ to $64^3$ were used.
The results shown below were obtained for the lattice size $32^3$
(the changes in the observables, as the lattice size is further
increased, are insignificant).
Once the field values on the lattice were determined, these were
transformed into their counterparts in the continuum theory
to obtain the statistical average value of the Hamiltonian
$\langle H\rangle$, i.e. the grand canonical (thermodynamical)
potential  $\Omega$ as a function of density.
%The equation of state - pressure versus density -
%is obtained from the relation $P = -\Omega_{\alpha}/V$,
%where $V$ is the  volume.
The energy of the  \al  matter per particle $\Omega_{\alpha}/N_{\alpha}$
is shown in Fig.~1 for the three values of the scattering
length $a_{\rm sc}^{(2)} = -7.33,$ -10, and -12 fm and the same
three-body interaction. For comparison we show also the zero temperature
variational results of Ref. \cite{CLARK4}. As expected, the thermodynamical
potential increases with the scattering length,
since the two-body potential scales linearly with $a_{\rm sc}$.
The close agreement of our result for $a_{\rm sc}= -7.33$~fm derived
from the VP-potential with the results of Johnson and Clark obtained
for the same interaction suggests that
at low temperatures neither the temperature dependence of the thermodynamic
potential $\Omega_{ \alpha}/N_{ \alpha}$, nor the contribution from
$l>0$ partial wave are significant
(taking the limit $T\to 0$ in the present theory is
complicated by the appearance of the Bose-Einstein condensate).
Our result for the larger scattering length $a_{\rm sc} = -12$
fm fit quite well the results of Johnson and Clark
based on the Ali-Bodmer (AB) potential \cite{AB}.

The critical temperature $T_c$ of the Bose-Einstein condensation
of \al particles can be obtained from the simulations. In practice we
obtain the function $n_{\alpha}(T)$ rather than directly $T(n_{\alpha})$
at constant fugacity $z\to 1$. Since we work at small
${\rm  log}~z= -0.1$, we compute the density of the system at constant
small chemical potential and temperature which is identified with
$T_c$. In this manner we extract an approximate
value of the critical temperature as a function of \al
matter density. Fig. 2 shows the density
dependence of the critical temperature of Bose-Einstein condensation
of \al matter extracted from simulations for two values of the
scattering length.
These temperatures are above the critical temperature
of the pairing in the $^3S_1$-$^3D_1$ pairing in a non-interacting
nucleon matter \cite{LOMBARDO}. While at low densities the  \al
condensation is the dominant effect, it should be kept in mind that
the \al particles lose their identity at higher densities when
their wave-functions strongly overlap and the Pauli-principle starts
to work for the constituent nucleons. The critical density for the
extinction of \al particles is sensitive to the magnitude of
the spatial extension of their wave-functions ($n_{\alpha}^*\propto
R_{\alpha}^{-3}$) and is of the order of $n_{\alpha}^*\simeq  0.03$ fm$^{-3}$ for
an effective radius $R_{\alpha}\simeq 2$ fm.  At densities
above the so-called Mott density \al particles are destroyed due to
the Pauli principle acting among their constituent
nucleons~\cite{ROEPKE}.

\section{Summary}
The finite temperature theory of  \al  matter close to
the critical temperature of the Bose-Einstein condensation
has been reformulated as a
scalar Euclidean $\phi^6$ theory with a negative quartic and positive
sextic interactions.
The theory is restricted to the zeroth order Matsubara modes
and is described by a real two-component field which is
$O(2)$ symmetric in the order parameter space.
The Monte-Carlo simulations of the theory on the lattice
established numerical relations between the
thermodynamic parameters (we work at constant
fugacity $z\approx 1$ and extract the density, the energy density
and the critical temperature).  The simulations are carried
out within an effective field theory with a cut-off, which
satisfies the condition $a_L\ll \xi$, where $\xi$ is
the correlation length. The obtained equation
of state is in a good agreement with the previous variational
calculations of the binding energy of infinite  \al  matter at
zero temperature \cite{CLARK1,CLARK2,CLARK3,CLARK4} in the
intermediate density regime. The equation of state of
low density alpha matter is presented here.

The critical temperature of  Bose-Einstein condensation of  \al
particles is found larger than the temperature for BCS condensation in
the isoscalar $^3S_1$-$^3D_1$ (quasi-deutron) channel. The dominance
of \al condensation is a low density phenomenon, since at densities
above the so-called Mott density \al particles are destroyed. The
mechanism is the Pauli principle acting among their constituent
nucleons~\cite{ROEPKE}. A treatment of the Mott effect requires
the knowledge of the internal structure of
the alpha particles and is  beyond the scope of present formalism.
Thus at asymptotically large densities (of the order of nuclear
saturation density) one is left with a quasi-deutron BCS
condensate \cite{LOMBARDO}.

 This work concentrated on a system consisting exclusively 
 of $\alpha$ particles. However, the Hamiltonian (\ref{eq:1}),
 which contains two-body attractive interaction, may have a spectrum 
 of bound states. 
    It is known that  $^{12}$C (the three alpha state)
    is bound, therefore our  system can evolve
    into a gas of $^{12}$C particles, and even further to the 
    point where the matter consists of $^{56}$Fe - 
    the most stable nucleus. As pointed out in the Introduction, 
    calculations of the matter properties under supernova conditions
    show a substantial fraction of alphas present in the matter, 
    which motivates the present study. Nevertheless, it will be 
    interesting to study the spectrum of bound states of the 
    Hamiltonian (\ref{eq:1}) and, for example, 
    the peculiarities related to the spectrum of the three-body bound
    state (Efimov effect)~\cite{BRAATEN_HAMMER}.

%\acknowledgments{ }
We are grateful to  Prof. J. W. Clark for communicating to us
refs. \cite{CLARK3,CLARK4} and to Dr. Kurt Langfeld for
discussions. This work was in part supported by a grant from the
SFB 382 of the DFG.

\end{document}